\documentstyle[mnextra]{mn}

\input{epsf}

\def\Msun{M_{\odot}}
\def\etal{{\it et al} \ }
\def\kms{km s$^{-1}$}
\def\gsim{ \lower .75ex \hbox{$\sim$} \llap{\raise .27ex \hbox{$>$}} }
\def\lsim{ \lower .75ex \hbox{$\sim$} \llap{\raise .27ex \hbox{$<$}} }
\def\Mpch{\, h^{-1}{\rm Mpc}}    
\def\kpch{\, h^{-1}{\rm kpc}}    
\def\kms{\, {{\rm km}}\,{{\rm s}}^{-1} }

\def\be{\begin{equation}}
\def\ee{\end{equation}}
\def\ia{\'{\i}}     

\pagerange{1--10}    
\pubyear{1998}
\volume{0}
\nobrackets
\seceqnum	
	
\begin{document}

\title[Galaxy destruction in clusters]
{Galaxy destruction and diffuse light in clusters}
\author[Calc\'{a}neo \etal]{Carlos Calc\'{a}neo--Rold\'{a}n$^1$\footnotemark, 
Ben Moore$^1$, Joss Bland--Hawthorn$^2$, David Malin$^2$ \\ \\ 
{\LARGE and Elaine M. Sadler.$^3$} \\ \\ 
$^1$Department of Physics, University of Durham, Durham DH1 3LE, UK.\\
$^2$Anglo-Australian Observatory, Epping Laboratory, P.O. BOX 296, Epping NSW. \\
$^3$University of Sydney, School of Physics, Sydney NSW 2006, Australia.} 

\date{November 1999}
 
\maketitle  

\begin{abstract} 

Deep images of the Centaurus and Coma clusters reveal two spectacular arcs of
diffuse light that stretch for over 100 kpc, yet are just a few kpc wide. At a
surface brightness of $m_b\sim 27-28$th arcsec$^{-2}$, the Centaurus arc is
the most striking example known of structure in the diffuse light component of
a rich galaxy cluster. We use numerical simulations to show that the Centaurus
feature can be reproduced by the tidal debris of a spiral galaxy that has been
tidally disrupted by the gravitational potential of NGC 4709.  The surface
brightness and narrow dimensions of the diffuse light suggest that the disk
was co-rotating with its orbital path past pericenter.  Features this
prominent in clusters will be relatively rare, although at fainter surface
brightness levels the diffuse light will reveal a wealth of structure.  Deeper
imaging surveys may be able to trace this feature for several times its
presently observed extent and somewhere along the tidal debris, a fraction of
the original stellar component of the disk will remain bound, but transformed
into a faint spheroidal galaxy. It should be possible to confirm the galactic
origin of the Centaurus arc by observing planetary nebulae along its length
with redshifts close to that of NGC 4709.

\end{abstract}

\begin{keywords}
galaxies: evolution -- galaxies: clusters -- galaxies: interactions -- 
galaxies: formation.
\end{keywords}

\footnotetext{Email:C.A.Calcaneo-Roldan@durham.ac.uk}

\section{Introduction}
\label{s:Intro}

Detecting diffuse light in clusters has an enigmatic history spanning
several decades (de Vaucouleurs, 1960; Frei \etal 1994; Thuan \& 
Kormendy 1977; Uson \etal 1991; V\ia lchez--G\'{o}mez \etal 1994;
Bernstein \etal 1995; Tyson \etal 1995).  Using either CCDs or
photographic imaging, these observations have been plagued by
background subtraction, stray light within the telescope and optics,
and atmospheric scattering. This has made a quantitative analysis
difficult: the total amount of diffuse light, its colour, or its
radial distribution have not yet been accurately measured. These
techniques have lead to claims that as much as 70\% of the light
attached to galaxies may lie in a diffuse component.  More recently,
individual planetary nebulae have been detected, inbetween cluster
galaxies and with redshifts and velocities that place them inside the
cluster potential (Arnaboldi \etal 1996; Theuns \& Warren 1996;
Feldmeier \etal 1997).  Deep HST images of the Virgo cluster have also
revealed a large population of freely orbiting, red-giant stars
(Ferguson, Tanvir \& Hippel 1998).  These studies also indicate large
quantities of diffuse light exist in clusters.

Intra-galactic stars must have formed within galaxies and have been 
subsequently ripped out by gravitational tidal forces.  Mergers and slow tidal
interactions between galaxies are a well studied phenomenon that can produce
dramatic tidal tails of stellar debris ({\it c.f.} Toomre 1964, 
Barnes \& Hernquist 1992 and references within).  Analysis of dark matter 
halos within a galaxy cluster that formed hierarchically has demonstrated that 
mergers are very rare within rich virialised environments (Ghigna \etal 1998). 
However, the impulsive and resonant tidal shocks from rapid fly-by encounters 
between galaxies can also create tidal debris. The cumulative effect of these 
encounters can cause a dramatic morphological transition between Sc-Sd spirals 
to dwarf ellipticals/spheroidals (Moore \etal 1996, Rakos \etal 1997), 
whereas low surface brightness galaxies, with lower central
densities, can be completely disrupted leading to a possible origin of the
diffuse intra-cluster light (Moore \etal 1999).  This process has been named
``galaxy harassment'', and extends previous work on slow interactions between 
galaxies into the impulsive tidal processes that operate in galaxy clusters 
(Merrit 1985, Valluri 1993, Henriksen \& Byrd 1996, Moore \etal 1998, Dubinski 
1998).

In the absence of further perturbations, stars that are tidally removed from
galaxies will orbit in narrow streams that trace the orbital path of the
galaxy. In a cluster, the star streams will be subsequently heated and mixed on
a time-scale of a few crossing times, {\it i.e.} several billion years.  We
might therefore expect to find prominent features in the intra-cluster light
component from recently disrupted galaxies that have accreted into clusters a
few billion years ago.  However, with only a couple of documented examples, why
are prominent features as bright as these so rare?

The properties of the diffuse light, including its quantity, radial
distribution, clumpiness and colour, are of great interest for many
reasons.  As well as constraining the importance of gravitational
interactions as a mechanism for morphological transformation, we have
the possibility of using thousands of freely orbiting stars for
studying the cluster potential. Understanding the orbital biases of
stripped stars and their subsequent evolution within a clumpy
potential, will be vital in the interpretation of these velocity data.

Recently, Trentham \& Mobasher (1998) detected a low surface
brightness feature $\sim 80$ kpc long within the Coma cluster, that
may be the result of a high speed encounter between two galaxies.
Conselice \& Gallagher (1998) also find a wealth of fine scale
substructure and faint tidal features in a survey of several nearby
clusters. Here we ``re-discover'' a much more spectacular arc of diffuse 
light that stretches for over 100 kpc near NGC4709 within the Centaurus
cluster.  The stacked sequence of photographic images by David Malin
were first reported very briefly in the Anglo-Australian newsletter by
John Lucey over 16 years ago; no further attention has since been
given to these data. Using the same techniques we have also discovered
a second feature that lies near the centre of the Coma cluster that is
morphologically similar to the Centaurus arc.

This paper is organized as follows. Section 2 presents details 
of the two newly discovered arcs, including photographic and CCD
colour information of the Centaurus arc, ending with a discussion of
alternative mechanisms to tides for the origin of these
features. Section 3 is an attempt to reproduce the properties of the
Centaurus arc, using numerical simulations to follow the 
disruption of galaxies within a cluster potential. 
We summarise these results in Section 4.

\section{The Images}
\label{s:images}

The Centaurus arc was originally discovered by applying a photographic
amplification technique (Malin 1978) to three plates taken by Malcolm
Smith at the f/2.66 prime focus of the 4m telescope of the Cerro
Tololo Inter-American observatory (CTIO, 1974). The photographic emulsion
was Eastman Kodak type IIIaJ, hyper-sensitized by baking in nitrogen
before use.  Photographically amplified positive derivatives from
these plates were combined into one image (Malin 1981) to improve the
image quality and minimize processing non-uniformities. The arc was
clearly visible on each of the three copies, and its reality was
latter confirmed by photographic amplification of IIIaJ plates taken
with the 3.9 m Anglo-Australian Telescope and the 1.2 m UK Schmidt
telescope.

More recent CCD observations reveal that if this structure lies in the
cluster, it is $\sim 120 \kpch$ ($\sim 12$ arcmin) long and only
$1\--2\kpch$ ($\sim 10-15$ arcsec) wide (Throughout this paper 
$H_{0} = 100h {\rm Mpc} \kms$, $h=1$). The arc has very low surface brightness
($\mu_{B} \gsim 27.8$ mag arcsec$^{-2}$), is red in colour and points
towards the active elliptical galaxy NGC4696. The arc's colour
strongly suggests that it is made of stars, so its narrowness is
remarkable.  The arc is not perfectly straight and has a small
curvature along its length.

\begin{figure*}
\centering
\epsfxsize=\hsize\epsffile{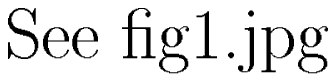}
\caption{A normal contrast image of the core of the Centaurus galaxy cluster 
with NGC 4696, the brightest galaxy in the cluster, at the upper right (lower 
panel) and a high-contrast version, with the extremely faint jet-like feature 
extending towards the lower left corner (upper panel). One arcmin at the 
distance of Centaurus is approximately $10\kpch$.}
\label{f:image}
\end{figure*}

The top half of Figure~\ref{f:image} shows a negative print of part of the 
photographically amplified, combined images of the CTIO plates. The arc 
is the linear feature that extends from the lower left corner (south east) 
towards the nucleus of NGC4696. The lower image shows the same part of the sky 
on a single unamplified plate. 

The photographs (and CCD frames) show the arc to be diffuse and seemingly 
devoid of fine structure at the arcsec level. While there are many faint 
stars and galaxies in the field, there is no apparent enhancement of 
point\--li\-ke or diffuse objects along its length. (The point\--li\-ke source 
near the centre of the arc is a star). The arc first becomes visible near a 
small, edge-on S0 galaxy, ESO 322-G102, at a projected distance of about 
$80\kpch$ from NCG 4696; the truncation  of the arc at this point may be a 
line-of-sight coincidence, since there is no evidence of any interaction 
between the arc and ESO 322-G102. Subtraction of the extended light profile of 
NGC 4696 may reveal the arc on the opposite side of the S0 galaxy.

Spectroscopy of the arc would be extremely difficult in view of its very low 
surface brightness, however, broadband CCD images of the brightest regions 
were obtained making it possible to compare the colours of the arc with aged 
stellar populations.  CCD pictures were taken in B, R and I bands with an RCA 
350x512 chip at the f/3.3 prime focus of the AAT under photometric conditions 
on the night of 21/22 June 1990. The CCD scale was $0.49$ arcsec/pixel, and 
the seeing $2-3$ arcsec. Two sets of overlapping exposures were taken, with 
total exposure times of 90 minutes in B, 40 mi\-nu\-tes in R and 20 minutes in 
I. Flat fields and bias frames were taken on the same night.

Table~\ref{t:one} lists the surface brightness and colours of the arc as 
measured from the overlap region of the CCD frames. In each case, the mean 
surface brightness in three regions along the arc was measured, each roughly 
$5\times5$ arcsec$^2$ and free of obvious foreground stars, and six `sky' 
regions of similar area straddling the arc and just outside it. The errors 
quoted in Table~\ref{t:one} are $1\sigma$ errors on the mean of the three 
sky-subtracted arc measurements in each filter.

\begin{table}
\begin{center}
\begin{tabular}{cc}
\hline
 Surface brightness (mag arcsec$^{-2}$) & Colour (mag) \\ \hline
\multicolumn {2}{c}{Arc} \\ \hline
 $\mu_B = 27.81 \pm 0.08$ & $B-R = +1.72 \pm 0.13 $ \\ 
 $\mu_R = 26.09 \pm 0.05$ & $R-I = +0.35 \pm 0.22 $ \\ 
 $\mu_I = 25.74 \pm 0.17$ &  \\ \hline
 \multicolumn {2}{c}{Sky (I frame in twilight)} \\ \hline
$\mu_B = 22.52 $ &   \\ 
$\mu_R = 20.76 $ &  \\
$\mu_I = 18.78 $ &   \\ \hline
\end{tabular}
\caption{Surface brightness and colours of the arc and sky.} 
\label{t:one}
\end{center}
\end{table}

The CCD measurements confirm that the arc is extremely diffuse and very faint, 
reaching no more than 0.7\% of the brightness of the night sky. Further out, 
the arc is even fainter and we estimate that the faintest parts of the 
structure revealed by the photographic plates are only 0.1\% of the night 
sky brightness. 

\begin{figure}
\centering
\epsfxsize=\hsize\epsffile{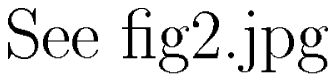}
\caption{A high-contrast image of the core of the Coma galaxy cluster with 
NGC 4874 at the upper right. One arcmin at the distance of Coma is 
approximately $30\kpch$.}
\label{f:image2} 
\end{figure}

The same techniques have also been applied to photographic images of
the central regions of the Coma cluster (Abell 1656).  These have
revealed a feature in the diffuse light, close to NGC 4874, that
stretches East-West for at least 5 arcmin, $\sim 150 \kpch$
(Figure~\ref{f:image2}). It is curved slightly concave to NGC 4874 in a
manner very similar to the curve in the Centaurus cluster feature where it
appears closest to NGC 4709. The image was made by combining
photographically amplified derivatives from three UK Schmidt
plates. Two of the plates (J9946 and J10027) were deep IIIa-J
(395-550nm) exposures while one was plate OR9945 covering the range
590-700nm. The linear feature is visible individually on all of the
plates, but is much less obvious on the red-light plate. Given the
large airmass through which the exposures were necessarily made and
the smaller number of plates, this suggests that the surface
brightness of the Coma arc is higher than that in the Centaurus
cluster. The large airmass has also contributed to the relatively poor
seeing in these plates, which is probably why we are unable to confirm
the Trentham \& Mobasher (1998) feature.

The Coma arc is neither as narrow nor as well defined as that in the
Centaurus cluster and two resolved galaxies appear to be embedded in
the brightest part of it. Given the large number of galaxies in the
field, this could be a line-of-sight coincidence, or one of these
could be the remnant nucleus of a disrupted galaxy. We note that this
feature in Coma was reported independently by Gregg \& West (1998). In the
absence of CCD photometry of the Coma arc, and its poorer resolution due
to its distance, we shall focus our attention on the origin of
the Centaurus arc.

\subsection{Possible origins}
\label{sub:origins}

The Centaurus arc is unlikely to be foreground reflection nebulosity in our 
own Galaxy. Malin has used his photographic amplifications technique on many 
fields containing Galactic nebulosity, and notes that the Centaurus feature 
(at Galactic latitude $22^{o}$) is morphologically quite different. In 
particular, it lacks the high-frequency `crumpling' characteristic of Galactic 
cirrus and reflection nebulosity. Also, the arc is almost straight (it 
deviates from a straight line by at most $3-4$ arcsec in the $100$ arcsec 
length covered by the CCD frames) and points at the nucleus of NGC 4696, the 
brigh\-test galaxy in the Centaurus cluster.

\begin{figure}
\centering
\epsfxsize=\hsize\epsffile{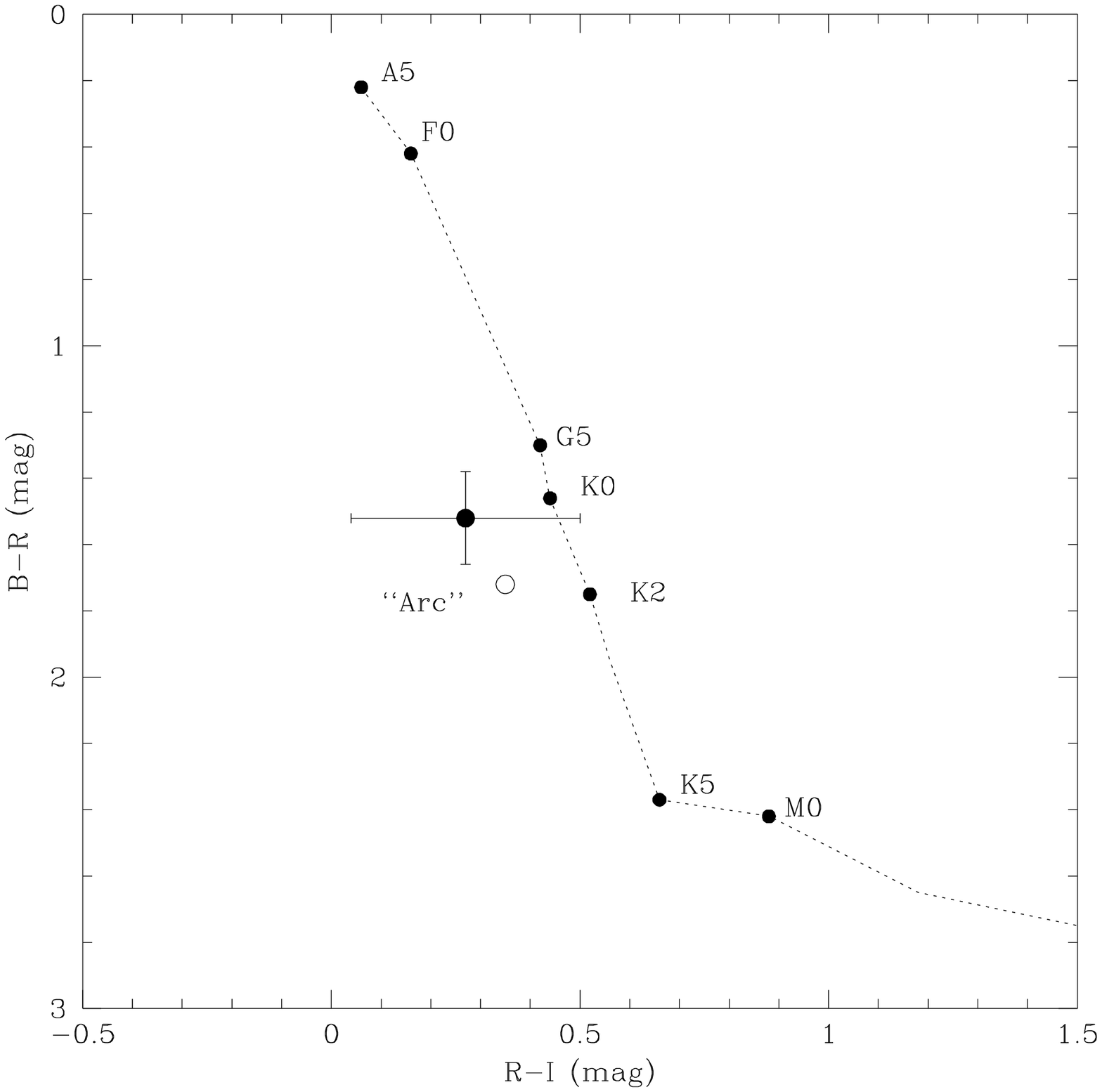}
\caption{The Colours of the Centaurus arc compared with those of late-type 
giant stars. The solid circle is the reddening corrected value, the open 
circle is the value without this correction.}
\label{f:kdata} 
\end{figure}

The region of the arc observed with the CCD has colours consistent with 
those of K0 stars in the (B-R), (R-I) two-colour diagram 
(Figure~\ref{f:kdata}, Cousins 1981). If the arc were dominated by optical 
synchrotron radiation it would be bluer than this, with B-R around 0.7-1.2 as 
typically seen in BL Lac objects (Moles \etal 1985) and the M87 jet 
(Tarenghi 1981). Whilst the arc might be composed of ionized gas, with most 
of its light coming from O$^{\scriptscriptstyle ++}$ and 
H$^{\scriptscriptstyle +}$ ions, very unusual 
line ratios would be needed to produce the observed colours and it would be
difficult to account for the I-band emission. Furthermore, it is hard
to imagine a long-lived ionizing source which could operate over such
a large distance. If light from the arc originated from emission lines, we
would also need to account for the co\-lli\-ma\-tion of the ionized gas, or
the ionizing beam, or both. We therefore conclude that the arc is
probably composed of stars with a mean spectral class of around K0.

Could the feature be a gravitational arc from a back\-ground galaxy
that has been lensed by the combined potential of NGC4696 and NGC4709?
In order to produce a gravitationally lensed image this straight, the
po\-ten\-tial has to be complex, such as would occur in between the
combined potential of the two central cDs. Furthermore, a lensed
image this close to the massive potential of NGC4709, would produce a
much shorter image.  Thus, if the mass distribution traces the light
distribution to a reasonable extent, then its position and morphology 
rule out gravitational lensing.

The dimensions of the object rule out a diffuse galaxy that happens to
lie along the line of sight \--- the axial ratios are about sixty to
one.  If the Centaurus arc is stellar and lies in the cluster, then
either the stars formed {\it in situ}, or they have been removed from
one of the cluster galaxies; since no mechanism is known for the
former, we shall concentrate on the latter.  In either case, the key
challenge for any successful model for its origin is to reproduce both
the length and narrowness of the feature.

We can estimate the mass of the stars in the arc from its integrated 
luminosity. Combining measurements of its area and mean surface brightness 
gives an estimated total B mag\-ni\-tude of $18.4 \pm 0.5$ for the integrated 
light. At the distance of the Centaurus cluster (taken here as $26.8\Mpch$), 
this is roughly $4h^{-2}\times10^7 L_\odot$, corresponding to 
$8h^{-2}\times10^7 M_\odot$ if we assume a mass-to-light ratio of 
${M}/{L_B} = 2$.

The remainder of this paper will be devoted to investigating the possibility
that the Centaurus arc consists of stars that have been tidally stripped from a
cluster galaxy.  Since the total stellar mass of the arc is just a few percent
of the stellar mass of an $L_*$ galaxy we have two possibilities for the
progenitor galaxy.  It may have originated from a single dwarf galaxy that has
been completely disrupted and all the original stars form the 100 kpc streak of
light. Alternatively, the observed feature may represent part of the tidal
debris that has been torn from a more luminous galaxy. In the latter case, the
bulk of the tidal debris may extend beyond the current image and may be
detectable at lower surface brightness levels.

\section{Tides and tails; numerical simulations}
\label{s:nbody}

We shall use numerical simulations to investigate the possibility that the
Centaurus arc is tidal debris from a gravitational interaction between a
galaxy and one of the cD galaxies NGC 4696 or NGC 4709.  Although their is a
wide parameter space to explore, in both morphology and orbits, it is clear
that spheroidal galaxies (either dwarf elliptical/dSph or giant ellipticals)
are unlikely candidates.  Ellipticals are too centrally concentrated to loose
a great deal of stars, and their tidal debris will not occupy such a narrow 
region in phase space. 

The position of the Centaurus arc next to the
cluster centre suggests that the potential of one of the massive
central cD galaxies was responsible for the disruption. However, we
can't rule out the possibility that the encounter took place further
from the cluster centre and we are observing the debris passing
pericentre.  Rather than treat the full cluster potential and its
substructure, we shall model the cD galaxy as a single truncated
isothermal dark matter halo with a moderate core radius of 50 kpc 
and mean velocity dispersion of $\sim 900$ kms$^{-1}$.

\begin{figure}
\centering
\epsfxsize=\hsize\epsffile{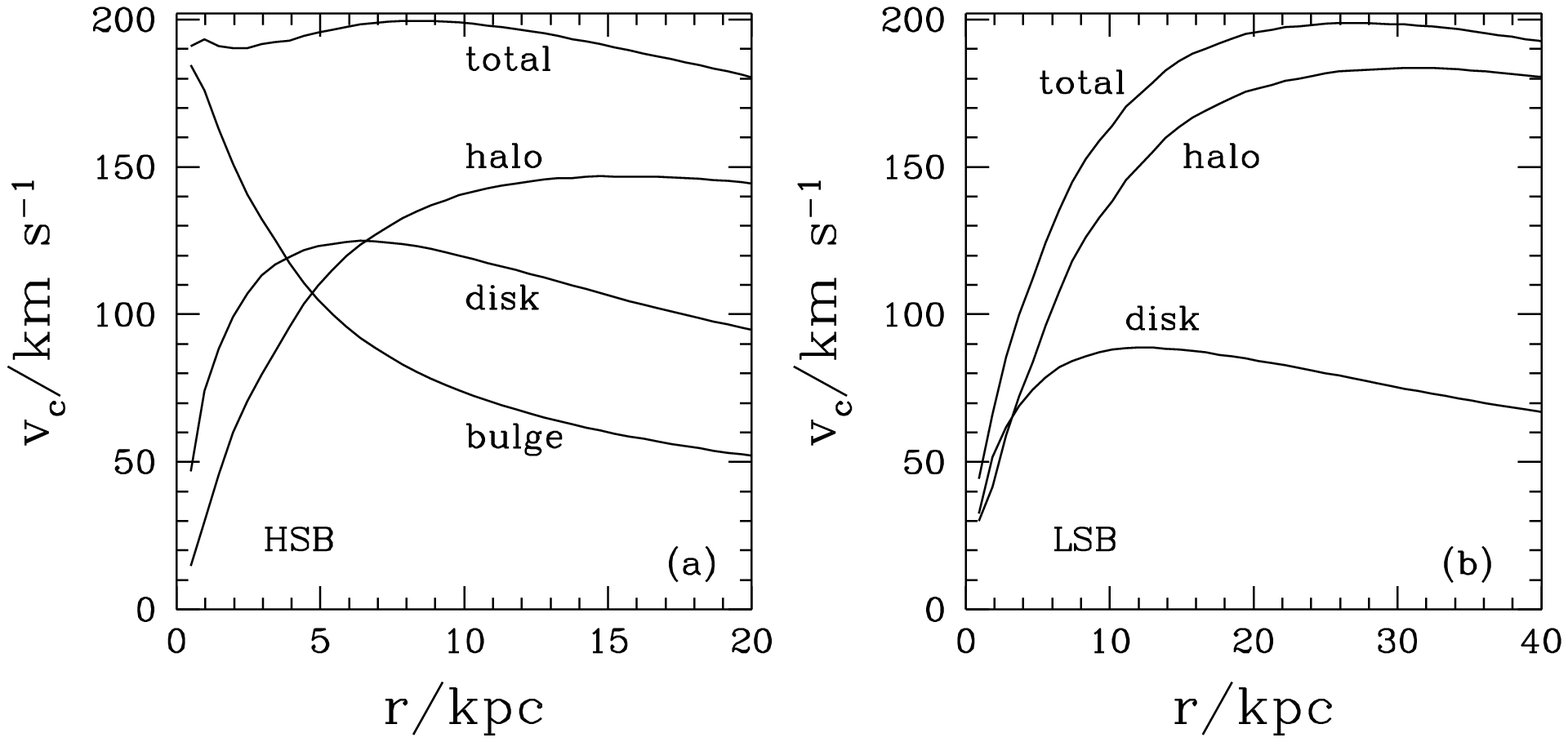}
\caption{The rotation curves $\propto \sqrt{M/r}$ of our model spiral galaxies with 
(a) high surface brightnes  
and (b) low surface brightness. The contributions to the rotation curve from the 
different components of the spiral galaxies are indicated.
Note that these models are constructed to represent observed galactic systems --
both have the same peak rotational velocity and lie on the same part of the Tully-Fisher
relation, yet have different central mass distributions.
}
\label{f:models}
\end{figure}

We will study galaxies with global properties that resemble
observed spheroidals and disks of varying surface brightness and luminosities
(c.f. Moore \etal 1999 and Figure 4.).  
Equilibrium ``N-body'' galaxy models are constructed
using the techniques of Hernquist (1989). To evolve the galaxies in orbit
through the cluster potential we use the parallel treecode ``PKDGRAV'' (Stadel
\etal , in preparation). When we simulate the galaxies in isolation they are
stable and remain in equilibrium.  We explore a wide range of orbits that
allow us to vary the strength of the impulsive shock and then compare the 
properties of the tidal debris with the Centaurus data.
Following the galaxies for several Gyrs we 
conclude that orbits with apocentric distance $\sim
1000$ kpc and pericentric distance of $\sim 150$ kpc produce the longest and
thinnest tidal streams.  This orbit is close to the typical orbit of ``halos
within halos'' for a hierarchical Universe measure by Ghigna \etal (1998).

For the spheroidal models, long streams of debris are obtained, 
but they are over 3 magnitudes too faint to explain the 
Centaurus arc.  Even after several passages past pericentre, most
of the material remains bound to the galaxies. We would require
the entire system to be disrupted into a single smooth stream
of length $\sim 100$ kpc in order to explain the Centaurus feature. 
We were unable to achieve this.

\begin{figure}
\centering
\epsfxsize=\hsize\epsffile{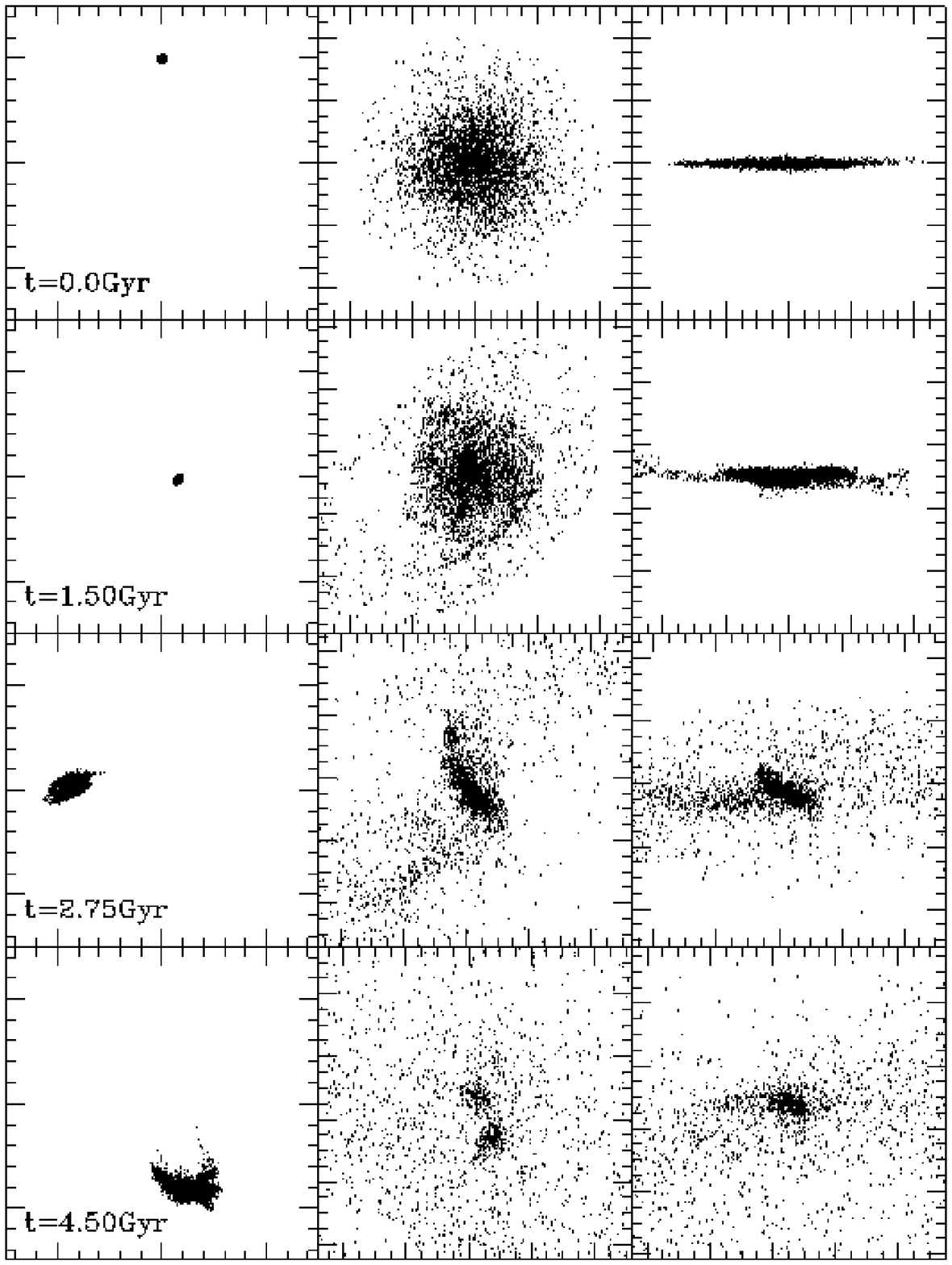}
\caption{The evolution of the LSB galaxy on a counter-rotating orbit.
The left panels show the entire orbit
centered on the cluster potential in a $3000^2$ kpc box. 
The two right hand panels measure $100$ kpc on the side and correspond to a 
close up view of the galaxy: face on, centre panel and edge on in the
right panel. Note that, for clarity, we plot just 1/5th of the star 
particles projected onto the orbital plane and the cluster particles are
not plotted to avoid confusion.}
\label{f:dotlc}
\end{figure}

The orbits of the spiral galaxies have an extra degree of freedom,
namely the orientation of the disk as the galaxy moves past
pericentre. We consider the extreme cases of a disk that is
either counter-rotating or co-rotating with respect to the orbital 
direction. Figure~\ref{f:dotlc} shows the evolution of one of the low 
surface brightness spirals in a counter-rotating orbit.

The first tidal shock occurs after 1.0 Gyr, yet after 1.5 Gyrs the
disk does not appear to be too disturbed. After 2.75 Gyr (almost 2.0 Gyrs 
after the encounter) the response to the perturbation is clearly apparent 
and the stripped stars surround the central stellar remnant. Continued 
heating after several more pericentric encounters almost completely 
unbinds the system, although the tidal debris never forms narrow 
features in phase space.

\begin{figure}
\centering
\epsfxsize=\hsize\epsffile{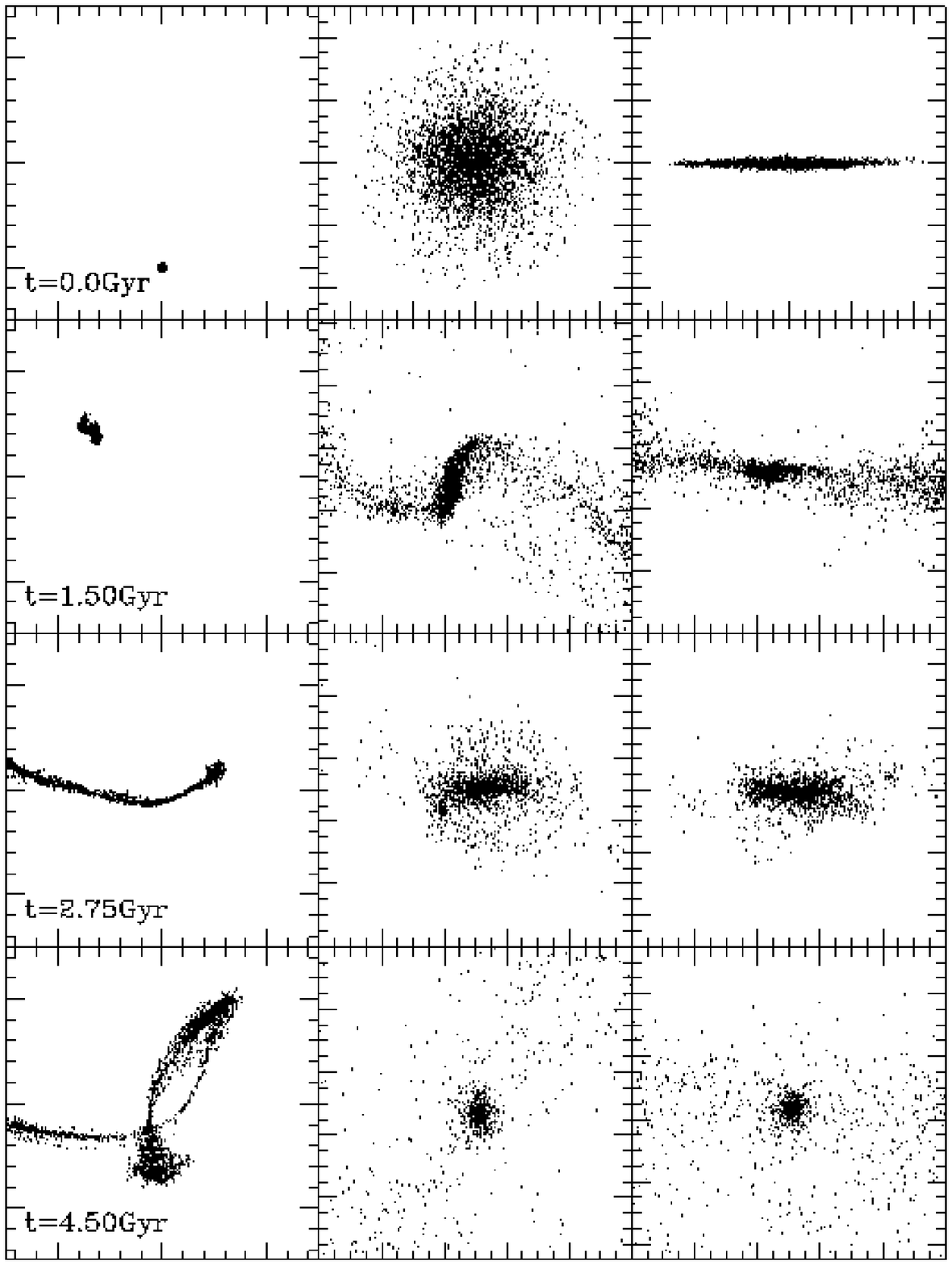}
\caption{As Figure ~\ref{f:dotlc}, except the LSB galaxy has been placed
on a co-rotating orbit.}
\label{f:dotl}
\end{figure}

We now change the direction of the orbit through the cluster potential
such that the disk is co-rotating with the galaxy's direction past
pericentre.  After just 0.5 Gyrs, the morphology of the galaxy
shown in Figure~\ref{f:dotl} has been dramatically altered.  Already, most
of the disk structure has been destroyed, and the stellar distribution
has been significantly heated.  However, the most significant change
occurs in the appearance of the tidal debris.  1.0 Gyr after the first 
encounter the tidal debris begins to form long thin tidal tails of stars 
that have been symmetrically torn from the disk and trace the orbital path 
of the galaxy. In this case, 90\% of the stars have been stripped, although
the remaining stars remain bound in a spheroidal configuration
with an exponential surface brightness distribution.

\begin{figure}
\centering
\epsfxsize=\hsize\epsffile{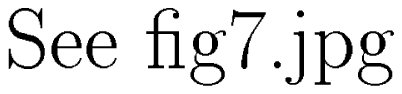}
\caption{
An illustration of the part of the tidal debris from the LSB disk galaxy that we
associate with the observed arc of light in the Centaurus cluster.  The star
particles from the LSB galaxy are plotted 2.75 Gyrs after the galaxy enters the
cluster, roughly 1.5 Gyrs after the first pericentric passage.  These data have
been inclined at $30^{\rm o}$ to the line of sight. At this time,
the stellar remnant is approaching pericenter for the second time. The small box
centred on the tidal debris shows the part of the stream that we may be
observing in the deep image of the Centaurus cluster shown above the tidal
debris.  The cross in the small box shows the centre of the cluster potential.
}
\label{f:overview}
\end{figure}

The tidal debris from both HSB and LSB galaxies can create long ($\gsim 200$ kpc) 
and thin ($\lsim 8$ kpc) diffuse arc-like features. 
Figure  ~\ref{f:dotl} shows that the debris has the narrowest
dimension and would be most luminous as it is passing pericentre. At
this point, the orbits of the stars bunch up because they are moving through
the deepest part of the potential. It is this section of the tidal debris
that we associate with the Centaurus arc that also lies
close to the centre of the cluster potential. We illustrate this in 
Figure~\ref{f:overview}.

Because the HSB galaxies are more centrally concentrated they lose less material
and the resulting tidal features are not as prominent.
Can we distinguish between these two possibilities? At time $t=2.75$ Gyrs
we extract a 100 kpc length of the stellar debris that is just approaching
pericentre and then project the data to measure its surface brightness.
The central surface brightness of the inner contour of the 
LSB galaxy is $\mu_B = 27.8$ arcsec$^{-2}$, while for the HSB it is 
$\mu_B = 28.3$ arcsec$^{-2}$. For this conversion we have assumed a mass to 
light ratio of ${M}/{L_B} = 2$. Although the difference is small, they
suggest that LSB galaxies produce brighter features although
a great deal of uncertainty arises from the assumed mass to light ratios.

A larger fraction of stars were stripped from
disks that are co-rotating with their orbit, and the stripped stars
formed long narrow streams that resembled the Centaurus arc.
We can understand why this happens by considering the relevant
dynamical time-scales.  The impulsive shock occurs on a time-scale
$t_o=2r_p/v_i$ Gyrs, where the impact velocity of the galaxy is
$v_i=3000\kms$ as it moves past pericentre $r_p=120$ kpc. We can
compare this time-scale with the time it takes for the disk stars to
make half a revolution within the core radius of the galaxy,
$r_{core}$, $t_i=\pi r_{core}/v_c$, where $v_c=200\kms$.  For the
particular galaxy and orbit simulated here, $t_i\sim t_o=0.1$ Gyrs.
If the disk is co-rotating, then the encounter occurs closer to 
a resonance and more energy is imparted to the disk stars, which 
can subsequently spread further through the cluster potential.

Our simulations showed that the orbits of the stripped stars move closer
together as they move through pericentre.  This creates the appearance of a
``standing wave'' near the cluster centre, where the surface brightness of the
debris is significantly enhanced.  The orbits bunch together near pericentre
because the gradient in the cluster potential is larger in the centre. We can
make a rough quantitative estimate of the enhancement in surface brightness as
follows: Consider two stars in circular orbits near the cluster centre at
distances r$_{a1}$ and r$_{a2}$, separated by a small radial distance
$\Delta$r$_a$.  What happens to the separation of the particles as we move the
orbits further out into the cluster, but preserve the small energy difference
between the two particles?  Now the particles orbit at distances r$_{p1}$ and
r$_{p2}$, this time separated by $\Delta$r$_p$.

The total energy of each orbit ($E_i$) is conserved and equal to 
$E_i = K_i + \Phi_i$ Where $K_i$ is the kinetic energy term for each orbit 
and $\Phi_i = 2 \sigma^2 ln \frac{r_i}{R}$\footnote{Note that we have use a 
{\it truncated} isothermal spherical potential where $\sigma$ is the constant 
velocity dispersion and $R$ is the truncation radius.} its corresponding 
potential energy. Because we are dealing with an isothermal potential, all 
circular 
orbits have the same kinetic energy, therefore the difference 
in total energy for each pair of orbits is given by:

\begin{center} 
$|\Delta E_A| = |\Phi_A| = |2 \sigma^2 ln ({r_{a2}}/{r_{a1}})| $ \\
$|\Delta E_P| = |\Phi_P| = |2 \sigma^2 ln ({r_{p2}}/{r_{p1}})| $
\end{center}
 
\noindent If the energy difference of both orbits is the same then

\begin{center}
$|\Delta E_A| = |\Delta E_B|$ \\
\medskip
$ ln ({r_{a2}}/{r_{a1}}) = ln ({r_{p2}}/{r_{p1}}) $
\end{center}

\noindent which leads to

\begin{center}
$ \Delta {\rm r}_a = ({{\rm r}_{a1}}/{{\rm r}_{p1}}) \Delta {\rm r}_p .$ 
\end{center}

\noindent Thus for a given energy difference; orbits tend to get closer 
together as they move towards the central regions of the potential. For
an orbit with apo:peri of 10:1, the enhancement of the surface brightness
the tidal stream will be roughly a factor of 10 at pericentre.

\begin{figure}
\centering
\epsfxsize=\hsize\epsffile{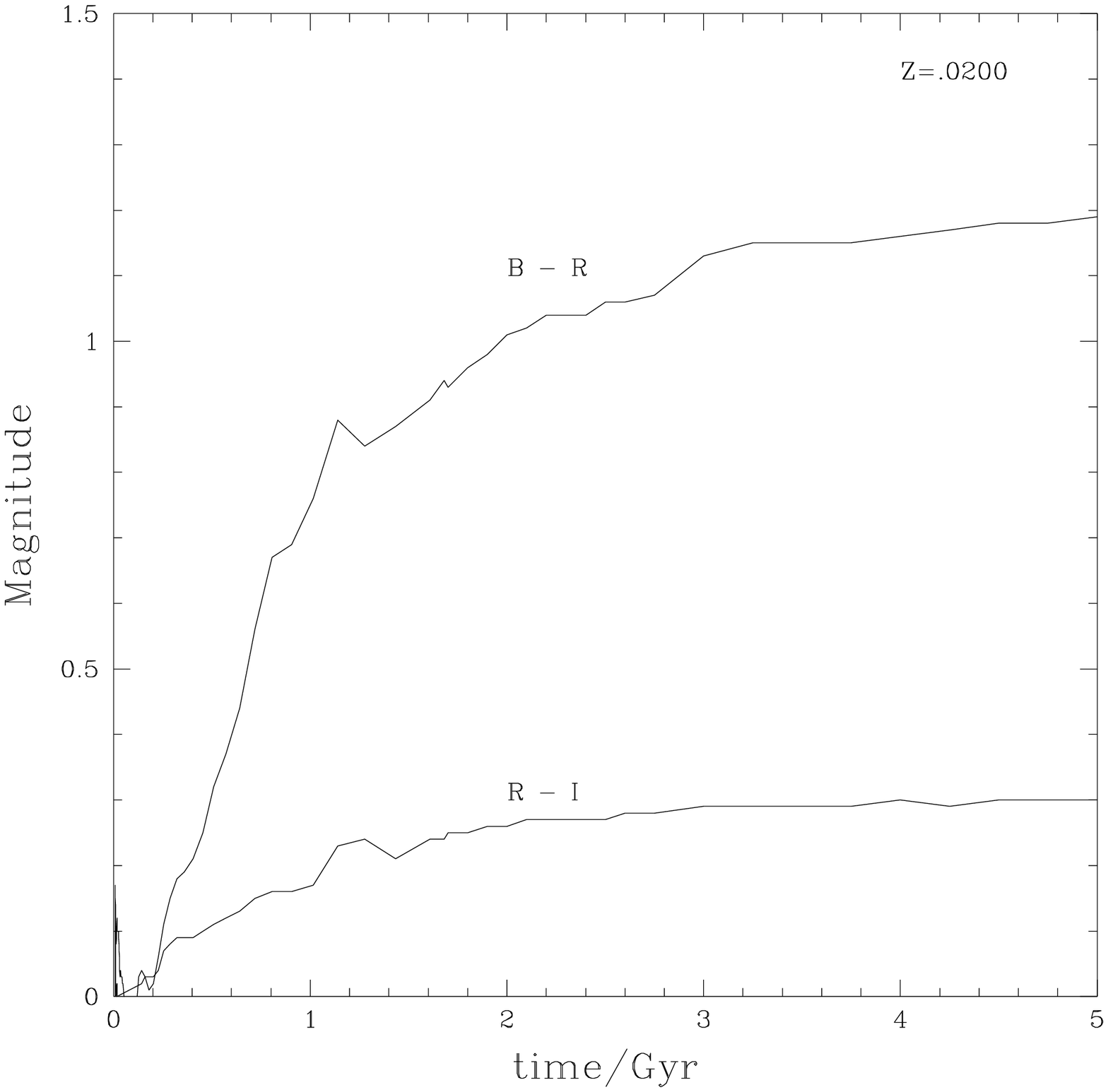}
\caption{Colour fading as a function of time for B-R and R-I colour 
differences.}
\label{f:mags}
\end{figure}

Further support for stellar debris of a galactic origin can be found
by considering the colours of the Centaurus arc.  In Table~\ref{t:two} 
we give typical colour differences for galaxies of
different morphologies across the Hubble diagram (Frei \& Gunn 1994;
de Blok \etal 1995). Once the stars are removed
from the galaxy, star formation will be abruptly halted and the stars
will fade in a predictable manner.  Figure~\ref{f:mags} shows
how the colour indices fade with time for a given stellar population
with known metallicity and IMF (Bruzual \& Charlot).

The tidal tails match the appearance of the Centaurus arc 
$\sim 1 $Gyr after being stripped from the model galaxies. i.e. the time since the
first passage past pericenter.
After a Gyr, the amount of fading will be $0.77$ and $0.17$
for B-R and R-I, respectively.  We now reconsider the observed values for
the arc (see Table~\ref{t:one}). We add a further correction for galactic dust 
reddening using the data of Burstein \& Heiles (1982) and Schlegel \etal 
(1998); this brings the values of the colours to:

\begin{center}
$B-R = 1.52 \pm 0.14$ \\
$R-I = 0.27 \pm 0.23$
\end{center}

\noindent 
If we take into account the amount of fading over one Gyr, then the
initial stellar colours of the stars in the arc would have been
$B-R=0.75$ and $R-I=0.10$. These colours, within the uncertainties, are 
consistent with late type spirals and LSB disk galaxies, providing further 
support for our model.

\begin{table}
\begin{center}
\begin{tabular}{ccccc}
\hline
 & Ellipticals  & Sab's & Scd's & LSB   \\ \hline
B-R  & $1.48$ & $ 1.04 $ & $ 0.86 $ & $ 0.78 $ \\ 
R-I  & $0.57$ & $ 0.57 $ & $ 0.43 $ & $ 0.49 $ \\ \hline 
\end{tabular}
\caption{Typical colour differences for different galaxy morphologies.} 
\label{t:two}
\end{center}
\end{table}

\section{Conclusions}
\label{s:concs}

Deep photographic and CCD observations of the Centaurus cluster
revealed a spectacular arc of diffuse light. This feature is
remarkable given its length and narrowness, $\sim 12$ arcmin $\equiv
120h^{-1}$ kpc long and $\sim 10$ arcsec $\equiv 2h^{-1}$ kpc wide. The
arc is diffuse with no apparent structure and its colours indicate
that it is made of stars. The estimated total mass from its integrated
luminosity is $\sim 8h^{-2}\times10^7 \Msun$ and its surface
brightness (in mag arcsec$^{-2}$) is $\mu_B = 27.8$; $\mu_R = 26.1$; in
the R band and $\mu_I = 25.7$, in the I band.  Several possible
scenarios for its origin, including foreground reflection nebulae,
gravitational lensing or a radio jet, are rejected in favor of a
gravitational tidal interaction that created an arc of stellar debris.
A second feature with similar morphology is also revealed within 
the central region of the Coma cluster.

We used numerical simulations to investigate the response of galaxies
of different morphologies to tidal shocks as they pass pericentre in a
cluster potential. The only scenario that gave rise to tidal debris with the
same characteristics of the Centaurus arc was a luminous spiral
galaxy with a disk co-rotating with its passage past pericenter.
This encounter geometry imparts the maximum energy
to the disk stars allowing them to stream away 
and form long thin tidal tails of stellar debris that trace
the orbital path of the galaxy. Only a small fraction of the tidal debris 
constitutes the Centaurus feature, which is prominent at its current
pericentric position where the orbits move closer together.

One could potentially confirm the galactic origin of the Centaurus
arc. By taking images along its length using different filters, ({\it e.g.}
OIII or H$\alpha$) such as discussed in Feldmeier \etal (1997), one would expect to
find an over abundance of planetary nebulae at similar redshifts to NGC4709;
thus confirming the stellar nature and formation mechanism.  Deeper images of
these features should allow them to be traced to larger extents. Somewhere
along the tidal tails lies the remnant spheroidal galaxy surrounded by a cloud
of diffuse light that closely resembles the feature reported by Trentham \&
Mobasher (1998).

\section*{Acknowledgments}

Carlos Calc\'{a}neo--Rold\'{a}n would like to thank the People of
M\'{e}xico for there generous support, through a grant by CONACyT,
which allows him to continue his research. Ben Moore is a Royal
Society research fellow. Computations were carried out as part of the
Virgo consortium.

\end{document}